\documentclass[12pt,journal]{IEEEtran}
 
\newcommand{\ignore}[1]{}

\usepackage{amsmath, amsthm, amsfonts, amssymb} 

\usepackage{float}
\newfloat{figure}{H}{lof}
\floatname{figure}{\figurename}

\DeclareMathAlphabet{\eufrak}{U}{}{}{}  % Euler fraktur math
\SetMathAlphabet\eufrak{normal}{U}{euf}{m}{n}
\SetMathAlphabet\eufrak{bold}{U}{euf}{b}{n}

\usepackage{amsmath, amsthm, amsfonts, amssymb}

\numberwithin{equation}{section}

\def\real{{\mathord{{\rm I\kern-2.8pt R}}}}        % Fake blackboard bold R.
\def\inte{{\mathord{{\rm I\kern-2.8pt N}}}}
\def\PP{{\mathord{{\rm I\kern-2.8pt P}}}}

\def\real{{\mathord{\mathbb R}}}

\def\inte{{\mathord{\mathbb N}}}

\newcommand{\disp}{\displaystyle}

\def\P{\mathbb{P}}
\def\E{\mathop{\hbox{\rm I\kern-0.20em E}}\nolimits}

\newtheorem{prop}{Proposition}[section]

\newtheorem{lemma}[prop]{Lemma}
\newtheorem{definition}[prop]{Definition}
\newtheorem{corollary}[prop]{Corollary}
\newtheorem{theorem}[prop]{Theorem}
\newtheorem{remark}[prop]{Remark}
\newtheorem{remarks}[prop]{Remarks}

\author{
Anthony R\'eveillac\footnote{areveill@mathematik.hu-berlin.de}  
\\ 
Institut f\"ur Mathematik
\\ 
Humboldt-Universit\"at zu Berlin 
\\ 
Unter den Linden 6 
\\ 
10099 Berlin 
\\ 
Germany
}

\allowdisplaybreaks 

\begin{document}

\title{Likelihood ratios and Bayesian inference for Poisson channels
} 

\hyphenation{func-tio-nals} 
\hyphenation{para-me-ter}

\maketitle

\begin{abstract}
\noindent 
In recent years, infinite-dimensional methods have been introduced for the Gaussian channels estimation. The aim of this paper is to study the application of similar methods to Poisson channels. In particular we compute the Bayesian estimator of a Poisson channel using the likelihood ratio and the discrete Malliavin gradient. This algorithm is suitable for numerical implementation via the Monte-Carlo scheme. As an application we provide an new proof of the formula obtained recently in \cite{GuoShamaiVerduPoisson} relating some derivatives of the input-output mutual information of a time-continuous Poisson channel and the conditional mean estimator of the input. These results are then extended to mixed Gaussian-Poisson channels.
\end{abstract} 
 
%\normalsize

%\vspace{0.5cm}


%\tableofcontents

%\normalsize

%\baselineskip0.7cm

\section{Introduction}
\label{section:Introduction}
Recently in \cite{Zakai}, infinite-dimensional methods have been used to derive a new expression of the conditional mean estimator for infinite-dimensional additive Gaussian channels. More precisely the conditional mean estimator is obtained as the Malliavin derivative of the logarithm of the likelihood ratio. In \cite{Zakai} this relation is used to show that the derivative of the input-output mutual information with respect to the signal-to-noise ratio of an additive Gaussian channel can be expressed in terms of the risk of the conditional mean estimator of the input. This fundamental connection has been first established in \cite{GuoShamaiVerduGaussian} using a different approach. In addition, the counterpart for Poisson channels of this connection has been obtained very recently by Guo, Shamai and Verd\'u in \cite{GuoShamaiVerduSA} and \cite{GuoShamaiVerduPoisson}. The aim of this paper is two-fold: first we prove that for a general Poisson channel, the conditional mean estimator can be obtained as the discrete Malliavin gradient of the likelihood ratio. Then, as an application, we present a new proof of the connection mentioned above obtained in \cite{GuoShamaiVerduSA} and \cite{GuoShamaiVerduPoisson}. Note that as an intermediate result, we provide extended de Bruijn identities (in the sense of \cite[Section VI]{Zakai}). Let us make more precise the statements mentioned previously.\\\\ 
In the general framework of \textit{additive Gaussian channel}, an observed signal $Y$ is decomposed into the sum of an \textit{input} signal $X$ plus an independent Gaussian noise $w$ as 
\begin{equation}
\label{eq:abstractadditive} 
Y=\rho \, X + w,
\end{equation}
where $\rho$ is the \textquotedblleft signal to noise ratio\textquotedblright. In this context the signals \textquotedblleft lie\textquotedblright\, in an \textit{abstract Wiener space} $(W,H,\mu_W)$ where $W$ is a separable Banach space, $H$ is an Hilbert space densely and continuously embedded in $W$ and $\mu_W$ is a Gaussian measure on $W$. In particular the input (resp. the output) $X$ (resp. $Y$) is an $H$-valued (resp. $W$-valued) random variable. This setting contains the case of an observed continuous-time stochastic process $(Y_t)_{t \in [0,T]}$ (with values into the space of continuous functions $W:=\mathcal{C}([0,T])$) related to an input stochastic process $(X_t)_{t \in [0,T]}$ (with values into the Hilbert space $H:=L^2([0,T])$) by the following stochastic differential equation,
\begin{equation}
\label{eq:CTGC}
dY_t = \rho \, X_t \, dt + dW_t, \; t\in [0,T]
\end{equation}
where $(W_t)_{t \in [0,T]}$ is a real valued standard Brownian motion independent of $(X_t)_{t \in [0,T]}$ and $\rho$ denotes the \textquotedblleft signal to noise ratio\textquotedblright. In \cite[Prop 4.1]{Zakai}, it is shown that
\begin{equation}
\label{eq:Zakai}
\E[X|\mathcal{Y}]=\frac{1}{\rho}\nabla \log l(Y),
\end{equation} 
where $\mathcal{Y}$ denotes the sigma field generated by $Y$, $\nabla$ denotes the Malliavin gradient which is a infinite-dimensional counterpart of the usual derivative on $\real^n$ and $l$ is the likelihood ratio associated to model (\ref{eq:abstractadditive}) that is, 
$$\disp{l:=\frac{d\mu_Y}{d\mu_W}}.$$ 
Relation (\ref{eq:Zakai}) entails the following result (\cite[Proposition 5.1]{Zakai}),
\begin{equation}
\label{fundamentalrelationGaussian}
\frac{dI(X;Y)}{d\rho} = \E\left[ \left\| X-\E[X\vert \mathcal{Y}] \right\|_H^2 \right],
\end{equation}
\textit{where $I(X;Y)$ denotes the mutual information between $X$ and $Y$,} defined as,
$$ I(X;Y):=\int_{H\times W} \log \frac{d\mu_{X,Y}}{d(\mu_X \times \mu_Y)} \mu_{X,Y}(dx,dy).$$
This relation had been previous obtained in \cite{GuoShamaiVerduGaussian} for time-continuous Gaussian channels using different techniques. Regarding these results one can ask the following question: can we find counterparts of relations (\ref{eq:Zakai}) and (\ref{fundamentalrelationGaussian}) in a non-Gaussian setting? An answer has been recently given in \cite{GuoShamaiVerduSA} and \cite{GuoShamaiVerduPoisson} for the Poisson regime. Let $Y=(Y_t)_{t\in [0,T]}$ be a Poisson process on $[0,T]$ with intensity measure $(X_t)_{t\in [0,T]}=\left(\int_0^t \lambda + \alpha \dot{X}_s ds\right)_{t \in [0,T]}$ where $\alpha, \lambda>0$ and $\dot{X}$ is a positive stochastic process. Then it is shown in \cite[Theorems 3-4]{GuoShamaiVerduPoisson} that
\begin{eqnarray}
\label{Theo4GSV}
&&\hspace{-1.8cm}\frac{d}{d\alpha} I(X;Y)\nonumber\\
&&\hspace{-1.8cm}=\frac{1}{\alpha}\E\left[ \int_0^T \psi_\lambda(\alpha \dot{X}_s+\lambda)-\psi_\lambda(\E[\alpha \dot{X}_s\vert\mathcal{Y}]) \nu(ds)\right]
\end{eqnarray}
and 
\begin{eqnarray}
\label{Theo3GSV}
&&\hspace{-1.8cm}\frac{d}{d\lambda} I(X;Y)\nonumber\\
&&\hspace{-1.8cm}=\E\left[ \int_S \log(\alpha \dot{X}_s+\lambda)-\log(\E[\alpha \dot{X}_s\vert\mathcal{Y}]) \nu(ds)\right].
\end{eqnarray}  
In this paper, we first extend Zakai's results to Poisson channels (see Proposition \ref{rapportBayesGrad} and Corollary \ref{corollary:rapportBayesGrad}). Then as an application, we provide a new proof of relations (\ref{Theo4GSV}) and (\ref{Theo3GSV}) in Theorem \ref{theorem:mutualPoisson}. As an intermediate result we also state and prove in Proposition \ref{DeBruijntypeidentities} extended De Bruijn identities analogous to \cite[Relation (35)]{Zakai}.\\\\
\noindent
We proceed as follows. First in Section \ref{section:classicalPoisson channel} we extend Relation (\ref{eq:Zakai}) to the setting of classical Poisson channels. Secondly, we will use infinite-dimensional stochastic analysis methods presented in Section \ref{section:Analysis on the Poisson space} to derive in Section \ref{section:cmePoisson} an equivalent of (\ref{eq:Zakai}) for infinite-dimensional Poisson channels using a Malliavin gradient for Poisson processes. This relation will be used in Section \ref{section:mutualPoisson} in order to give an new proof of (\ref{Theo4GSV}) and (\ref{Theo3GSV}) for general Poisson channels. Then in Section \ref{subsection:w+PCME}, we generalize the results obtained in Section \ref{section:cmePoisson} to a class of normal martingales which contains the continuous time Poisson channel, the Gaussian one and a mixture of the both (this class includes some martingales with jumps and non-independent increments). Finally in Section \ref{section:mutualmixture} we extend relations (\ref{Theo4GSV}) and (\ref{Theo3GSV}) to a deterministic mixture of Gaussian and Poisson channels. We remark that we were not able to show relations of the type (\ref{Theo4GSV}) and (\ref{Theo3GSV}) for the processes with non-independent increments presented in Section \ref{subsection:w+PCME} Example 2). This phenomenom was suggested in the last sequence of \cite[Section VI]{GuoShamaiVerduPoisson} where it has been remarked that both Gaussian and Poisson processes share the independent increments property.

\section{Poisson channel on $\inte$}
\label{section:classicalPoisson channel}

Let us briefly describe the Poisson channel on $\inte$ (see \cite{Verdu} for a survey on Poisson channels).\\
\noindent 
Poisson channels are different from Gaussian channels in the sense that the observed signal cannot be expressed as the sum of the input signal plus some additional noise, it cannot be expressed in an \textquotedblleft additive\textquotedblright \, way like in (\ref{eq:abstractadditive}). Consider a positive input signal $X$ with distribution $\mu_X$. We assume the output $Y$ is a Poisson random variable on $\inte$ with intensity $\alpha X+\lambda$, 
$$ Y \sim \mathcal{P}(\alpha X+\lambda).$$
This setting is used for example in photo-detection problems where a photo-sensitive device (\textit{e.g.} a p-i-n diode) is modeled by a Poisson channel. In this setting $\lambda$ is a residual current in the device called the \textquotedblleft dark current noise\textquotedblright\, and $\alpha$ is some scale parameter. Note that contrary to the Gaussian channel $\lambda$ and $\alpha$ cannot be replaced by a single coefficient, the \textquotedblleft signal to noise ratio\textquotedblright. \\
Let $\mu_{0}$ be the distribution of a Poisson random variable on $\inte$ with intensity $1$. Finally assume that the conditional law $\mu_{Y\vert X}(\cdot\vert x)$ is absolutely continuous with respect to $\mu_{0}$ (this condition implies that the joint distribution of $(X,Y)$ is absolutely continuous with respect to measure $\mu_X \times \mu_{0}$) whose density is given by
\begin{eqnarray*}
&&\disp{\frac{d\mu_{Y|X=x}}{d\mu_{0}}(y)}\\
&=&\disp{\exp(-((\lambda-1)+\alpha x)) \left(\lambda+\alpha x\right)^y,}
\end{eqnarray*}
for $x\in \real_+, y \in \inte$, and the law of $Y$ is absolutely continuous with respect to $\mu_{0}$ with density $m$,
\begin{equation}
\label{eq:unconditionaldensityfinitedimensional}
m(y):=\int_{0}^{+\infty} \frac{d\mu_{Y|X=x}}{d\mu_{0}}(y) \, \mu_X(dx), \; y \in \inte. 
\end{equation}
Now we can state the following lemma which will be extended in Section \ref{section:cmePoisson} as Proposition \ref{rapportBayesGrad} and Corollary \ref{corollary:rapportBayesGrad}.
\begin{lemma}
\label{lemma:BayesianPoissonChannel}
The Bayesian estimator of $\lambda+\alpha X$ can be expressed as:
\begin{equation}
\label{eq:BPCP}
\E[X|\mathcal{Y}]=\frac{m(Y+1)-m(Y)}{\alpha m(Y)}-\frac{(\lambda-1)}{\alpha}.
\end{equation}
\end{lemma} 
\begin{proof}
Let $y$ in $\inte$.
\begin{eqnarray*}
&&\displaystyle{ m(y+1)-m(y) }\\
&=& \displaystyle{\int_0^{+\infty} L(y,x) (\lambda-1+\alpha x)  \mu_X(dx) } \\
&\overset{(\ast)}{=}& \displaystyle{ m(y) \int_0^{+\infty} (\lambda-1+\alpha x) \frac{d\mu_{X|Y=y}}{d\mu_X}(x) \, \mu_X(dx)}\\
&=& \displaystyle{m(y) \left(\lambda-1+\alpha \int_0^{+\infty} x \mu_{X|Y=y}(dx)\right)}\\
&=& \displaystyle{m(y)(\lambda-1+\alpha \E[X|Y=y]) }
\end{eqnarray*}
Equality $(\ast)$ is justified by a relation of the form $iii)$ of Proposition \ref{prop:BayesJust}.
\end{proof}
\begin{remark}
The nonlinear filter of $X$ given in (\ref{eq:BPCP}) can be numerically approximated thanks to a Monte-Carlo scheme (see Remark \ref{remark:application}).
\end{remark}
\begin{remark}
\label{rem:section2}
The conditional distributions used in Lemma \ref{lemma:BayesianPoissonChannel} are well defined in this context, one can refer to Propositions \ref{prop:BayesJust} and \ref{prop:BayesPoisson} for more details. 
\end{remark}
\noindent
To obtain results for more general Poisson channels we have first to recall some elements of analysis on the Poisson space.

\section{Analysis on the Poisson space}
\label{section:Analysis on the Poisson space}
 
In this Section we introduce some elements of analysis on the Poisson space in a general framework. We will then describe these elements using a concrete example.\\
Let $(S,\mathcal{B}(S),\nu)$ a measure space where $\nu$ is an intensity measure that is atomless and $\sigma$-finite. For an element $z$ in $S$ we denote by $\delta_z$ the Dirac-measure at point $z$ on $(S,\mathcal{B}(S))$. Define the Poisson space $\Omega_S$ as 
$$\Omega_S = \left\{y=\sum_{k=1}^n \delta_{z_k}, \; n \in \bar{\inte}, \; z_k \in S, \; 1 \leq k \leq n \right\},$$
with $\bar{\inte}:=\inte \cup \{ \infty \}$ and for $y=\sum_{k=1}^n \delta_{z_k}$, let 
\begin{equation}
\label{definition:JT}
\mathcal{C}(y):=\{z_1, \ldots, z_n \}.  
\end{equation}
Define the canonical process $(N_A)_{A\in \mathcal{B}(S)}$ on $\Omega_S$ as
$$N_A(y):=y(A), \; y \in \Omega_S,$$
where $y(A)$ is the number of atoms of $y$ in the set $A$. We define the $\sigma$-field $\mathcal{F_S}$ on $\Omega_S$ with $\mathcal{F_S}=\sigma(\{ y \mapsto y(B), \; B \in \mathcal{B}(S) \})$.\\
There exists a probability measure $\P_S$ on $(\Omega_S,\mathcal{F}_S)$ called the Poisson measure such that,
\begin{itemize}
\item $\displaystyle{ \forall B \in \mathcal{B}(S), \; \forall n \in \inte,}$\\  
$\displaystyle{ \P_S(\{y \, | \,y(B)=n \}) = \exp(-\nu(B)) \, \frac{\nu(B)^n}{n !} }$ 
\item For disjoint subsets $\displaystyle{ (B_1,\cdots,B_n) \textrm{ in } \mathcal{B}(S)}$,\\ 
$\displaystyle{y(B_1), \; \ldots, y(B_n) }$ are $\P_S$-independent. 
\end{itemize}
Under $\P_S$ the canonical process $(N_A)_{A\in \mathcal{B}(S)}$ is a Poisson process with intensity $\nu$.\\  
Define $\mathcal{M}(S)$ as the set of non-negative measure on $(S,\mathcal{B}(S))$. Let $H_S$ be the space 
\begin{eqnarray*}
&&H_S=\bigg\{ \omega \in \mathcal{M}(S), \exists  h \in L_+^2(S,d\nu),\\
&&\hspace{2cm}\omega << \nu \textrm { with } \frac{d\omega}{d\nu}=h \bigg\},
\end{eqnarray*}
where $L_+^2(S,d\nu)$ denotes the set of positive function of $L^2(S,d\nu)$.\\
$H_S$ is equipped with an inner product $\langle \cdot,\cdot \rangle_{H_S}$ given by
$$ \langle \omega_1,\omega_2 \rangle_{H_S} = \langle h_1,h_2\rangle_{L^2(S, d\nu)}, \; \omega_1 \in H_S, \; \omega_2 \in H_S.$$
Note also that we will denote by $\E$ the expectation with respect to the measure $\mu_X\times \mu_Y$, $\E_1$ the expectation with respect to $\mu_Y$ and $\E_0$ the expectation relative to $\P_S$.\\\\
\indent
The Malliavin operator $\nabla$ we introduce will be of interest in Sections \ref{section:cmePoisson} and \ref{section:mutualPoisson}.\\
Let $L^0(\Omega_S,\mathcal{F}_S,\P_S)$ be the space of measurable mappings from $(\Omega_S,\mathcal{F}_S,\P_S)$ to $\real$. Define first the operator $D$ by,
\begin{eqnarray*}
\displaystyle{L^0(\Omega_S,\mathcal{F}_S,\P_S)}&\to&\displaystyle{L^0(\Omega_S \! \times \! S, \mathcal{F}_S \! \otimes \! \mathcal{B}(S), \P_S \! \otimes \! \nu) }\\
\displaystyle{F} &\mapsto& \displaystyle{D_z F(y):= F(y + \delta_z) - F(y).}
\end{eqnarray*}
Technical justifications about the measurability of the previous map can be found in \cite{Wu} and references therein. We mention the following chain rule property for the Malliavin derivative $D$, that is, for every random variables $F$ and $G$ on $(\Omega_S, \mathcal{F}_S, \P_S)$ we have that
\begin{equation}
\label{chainrulePoisson}
D_s(FG)=F D_sG + G D_s F+ D_sF D_s G, \; s \in S.
\end{equation}
We also introduce the operator $I_1$ and the Malliavin integration by parts formula which will play an import role in Section \ref{section:mutualPoisson}. For a deterministic function $h:S\to \mathbb{R}$ we denote by $I_1(h)$ the stochastic integral of $h$ against the martingale $y-\nu$, \textit{i.e.}, 
$$ I_1(h):=\int_S h(s) (dy_s-\nu(ds)).$$
Note that the stochastic integral is defined pathwise (in the sense of Stieljes) as follows
$$ \int_S h(s) dy_s=\sum_{k=1}^n h(z_k), \; a.s$$
where $y(S)=n$ and $y=\sum_{k=1}^n \delta_{z_k}$.\\\\
\noindent
Let $h$ as above and let $F$ be a random variable on $(\Omega_S, \mathcal{F}_s, \P_s)$ we have that
\begin{equation}
\label{IBP}
\E_0[F I_1(h)]=\E_0\left[ \int_S D_s F h(s) \nu(ds) \right].
\end{equation}
Note that the proof of this formula is done for example in \cite{Privault}. We define the operator $\nabla$ which is an "integrated" version of $D$.
\begin{definition}
For $F:\Omega_S \to \real$ we define $\nabla F$ as the $H_S$-valued random variable 
$$ \nabla_A F := \int_A D_z F \, \nu(dz), \; A \in \mathcal{B}(S) .$$
\end{definition}   
\noindent
We conclude this section by mentioning that the setting described above contains the canonical Poisson space $\Omega_{[0,T]}$ as a particular case, where 
$(S,\mathcal{B}(S),\nu)=([0,T],\mathcal{B}([0,T]),d\pi)$ with $\pi$ being the Lebesgue measure on $[0,T]$ and
\begin{eqnarray*}
&&\Omega_{[0,T]}\\
&&:=\bigg\{ y=\sum_{k=1}^n \delta_{t_k}, \, n \in \bar{\inte}, \, 0\leq t_1<\ldots<t_n \leq T \bigg\}.
\end{eqnarray*}
In this case $\mathcal{C}(y)$ given by (\ref{definition:JT}) is the set of the jump times of the path $y$ and under $\P_{[0,T]}$, $(N_{[0,t]})_{t\in[0,T]}$ is a Poisson process with intensity $dt$, that is, the stochastic process $(N_t-t)_{t \in [0,T]}$ is a $\P_{[0,T]}$-martingale.\\
In this case $H_{[0,T]}$ can be defined in a more tractable way by,
\begin{eqnarray*}
\disp{H_{[0,T]}=}&\bigg\{&\disp{\omega:[0,T] \to \real, \exists  h \in L^2([0,T]), }\\
&&\omega << \pi \textrm { with } \frac{d\omega}{d\pi}=h \bigg\},			
\end{eqnarray*} 
equipped with 
$$ \langle \omega_1,\omega_2 \rangle_{H_{[0,T]}} := \langle h_1,h_2 \rangle_{L^2([0,T])}, \; \omega_1, \omega_2 \in H_{[0,T]}.$$
Finally in the case of the classical Poisson space the Malliavin derivative $\nabla$ can be expressed in a different way, for $F:\Omega_{[0,T]} \to \real$, $\nabla F$ is a $H_{[0,T]}$-valued random variable and 
$$ \nabla_{[0,t]} F:= \int_0^t D_s F \, \pi(ds), \; t \in [0,T]. $$

\section{Conditional mean estimators for Poisson channels}
\label{section:cmePoisson}
\subsection{Some general facts about the Bayesian framework}
\label{subsection:generalBayesianfacts}
We introduce in this Section the Bayesian framework and compute in Section \ref{ivB} the conditional mean estimator in the setting of Poisson point process.\\\\
\noindent
Let $X$ be an input signal with values in a space $(H,\sigma(H))$ with distribution $\mu_X$. Consider $(\Omega,\mathcal{F},\P)$ a probability space and assume the output $Y$ lies in $\Omega$. We make the following assumptions;
\begin{itemize}
\item[(H1)] For all $x$ in $H$, $\mu_{Y|X=x}$ (the distribution of $Y$ given $X=x$) is absolutely continuous with respect to $\P$ and we denote by $L$ the corresponding Radon-Nikodym density.
\item[(H2)] $L$ is $(\sigma(H) \otimes \mathcal{F})$-measurable.
\end{itemize} 
Then, the following function 
$$
\left.
\begin{array}{l}
H \times \mathcal{F} \to [0,1] \\
(x,B) \mapsto \mu_{Y|X=x}(B)
\end{array}
\right.
$$ 
is a transition probability in the sense of \cite[Definition III-2-1 p. ~69]{Neveu}. Moreover the joint distribution $\mu$ of $(X,Y)$ is a probability measure on $(H \times \Omega, \sigma(H) \otimes \mathcal{F})$ such that,
\begin{eqnarray}
\label{eq:transition1}
&&\disp{\mu(A \times B)}\\
&=&\disp{\int_A \mu_{Y|X=x}(B) \, \mu_X(dx), \; A\times B \in \sigma(H)\otimes \mathcal{F}}\nonumber.  
\end{eqnarray}
Denote by $M$ the marginal distribution of $\mu$ on $(\Omega,\mathcal{F})$ defined by,
\begin{equation}
M(B):=\mu(H \times B), \; B \in \mathcal{F}.
\end{equation}
Proposition \ref{prop:BayesJust} is mainly devoted to show the existence of the following transition probability  
\begin{equation}
\label{eq:BayesInv}
\left.
\begin{array}{l}
\Omega \times \sigma(H) \to [0,1] \\
(y,A) \mapsto \mu_{X|Y}(A\vert y)
\end{array}
\right.
\end{equation}
and that the couple $(M,(\mu_{X|Y}(\cdot\vert y))_{y \in \Omega})$ allows us to recover $\mu$ as
\begin{equation}
\label{eq:transition2}
\mu(A \times B)=\int_B \mu_{X|Y}(A\vert y) \, M(dy), \; A \times B \in \sigma(H)\otimes \mathcal{F}.  
\end{equation}
\begin{prop}
\label{prop:BayesJust}
If \emph{(H1)} and \emph{(H2)} are satisfied then
\begin{itemize}
\item[\textit{i)}] $\mu$ is absolutely continuous with respect to $\mu_X \times \P $ and the corresponding Girsanov-Radon-Nikodym density is $L$.
\item[\textit{ii)}] $M$ is absolutely continuous with respect to $\P$. Let $m$ be a version of $dM/d\P$.
\item[\textit{iii)}] For $M$ almost all $y$ in $\Omega$, $\mu_{X|Y=y}$ is absolutely continuous with respect to $\mu_X$ and for $y$ such that $m(y) \neq 0$, the Radon-Nikodym density is given by,
$$ \frac{d\mu_{X|Y=y}}{d\mu_X}(x) = \frac{L(y,x)}{m(y)}.$$ 
\item[\textit{iv)}] For a $(\sigma(H) \otimes \mathcal{F})$-measurable function $f: H \times \Omega \to \real$,
\begin{eqnarray*}
&&\displaystyle{\int_\Omega \int_H f(x,y) \mu_{X|Y=y}(dx) \, M(dy)}\\ 
&=&\displaystyle{\int_H \int_\Omega f(x,y) \mu_{Y|X=x}(dy) \, \mu_X(dx)}.
\end{eqnarray*}   
\end{itemize}
\end{prop}
\begin{proof}
See for example \cite[Section 4.2.1, p.~126]{Berger}, or \cite[Section A.3, p.~623-626]{LieseMiescke}.
\end{proof}
\noindent
Now we will make use of the general Bayesian framework described above. 
\subsection{General Poisson channels}
\label{ivB}
Let $(S,\mathcal{B}(S),\nu)$ and $H_S$ as in Section \ref{section:Analysis on the Poisson space}. We denote by $(\Omega_S,\mathcal{F}_S,\P_S)$ the Poisson space introduced in Section \ref{section:Analysis on the Poisson space} and assume that under the probability measure $\P_S$, the output process $Y$ is a Poisson point process with intensity measure $\nu$. Let in addition $\lambda$ and $\alpha$ be positive numbers. Let $X$ be the input random variable with values in $H_S$ such that $\int_H \int_S \dot{x}_z \nu(dz)\mu_X(dx)<\infty$. Then, by Girsanov theorem (see for example \cite[Theorem 3.1.1, p.~78]{Reiss}), $\mu_{Y|X}(\cdot \vert x)$  the conditional probability on $\Omega$ given $X=x$ is absolutely continuous with respect to $\P_S$ and the Girsanov-Radon-Nikodym density denoted $L$ is given by
\begin{eqnarray}
\label{eq:vrais}
\displaystyle{L(y,x)}&:=&\disp{\frac{d\mu_{Y|X}(\cdot\vert x)}{d\P_S}(y)}\\
&=&\displaystyle{\exp \left( -(\lambda-1) \nu(S) -\alpha \int_S \dot{x}_z \nu(dz) \right)}\nonumber\\ 
&\times&\disp{\prod_{k=1}^{y(S)} (\lambda+ \alpha \, \dot{x}(z_k)),}\nonumber
\end{eqnarray} 
where $y(S)=n$ and $y=\sum_{k=1}^n \delta_{z_k}$. In other words, under the probability measure $\mu_{Y\vert X}(\cdot \vert x)$, the stochastic process $Y$ is a Poisson process with intensity $\lambda+\alpha x$.

\begin{prop}
\label{prop:BayesPoisson}
Assume that hypotheses (H1) and (H2) are in force. Let $\mathcal{B}_A(Y):=\E[X(A)|\mathcal{Y}]$ for $A \in \mathcal{B}(S)$. Then
\begin{eqnarray}
\label{eq:BayesPoisson1} 
&&\mathcal{B}_A(Y)\\\nonumber
&&=\int_{H_S} x(A) \, \mu_{X|Y}(dx\vert y),\textrm{ for } M-\textrm{almost every } y.
\end{eqnarray}
\end{prop}
\begin{remark}
Note that the expression (\ref{eq:BayesPoisson1}) is theoretical and cannot be used in practice. In contradistinction, relation (\ref{eq:rapportBayesGrad}) obtained below enables a numerical approximation of the Bayesian estimator as mentioned in Remark \ref{remark:application}. 
\end{remark}
\noindent
In fact it is more tractable to estimate the densities rather than the intensity measures. So we denote by $\dot{X}$ the $L^2(S,d\nu)$ valued random variable associated to $X$. For $z$ $\in S$ (\ref{eq:BayesPoisson1}) can be rewritten as
\begin{equation}
\label{eq:BayesPoisson2} 
\dot{\mathcal{B}}_z(y)=\E[\dot{X}_z|Y=y]=\int_{H_S} \dot{x}_z \, \mu_{X|Y}(dx\vert y), \; M-a.e.
\end{equation}
\noindent
We can state the main result of this paper. It allows us to express the Bayesian estimator of the input as a discrete logarithmic Malliavin gradient of the likelihood ratio $m$. We recall that
\begin{equation}
\label{eq:m}
m(y) = \int_{H_S} L(y,x) \, \mu_X(dx), \; y \in \Omega_S.
\end{equation} 
\begin{prop}
\label{rapportBayesGrad}
Assume that hypotheses (H1) and (H2) are satisfied then for $M$-almost every $y$ we have that
\begin{equation}
\label{eq:rapportBayesGrad}
\E[X(A)|\mathcal{Y}]=\frac{\nabla_A m(y)}{\alpha \; m(y)}-\frac{(\lambda-1) \nu(A)}{\alpha}, \; \forall A \in \mathcal{B}(S).
\end{equation}
\end{prop}
\begin{proof}
For $y$ in $\Omega_S$ we set: $y(S)=n$, $y=\sum_{k=1}^n \delta_{z_k}$ and $\mathcal{C}(y)$ be the set defined by (\ref{definition:JT}). Let $z$ in $S$, we have that
\begin{eqnarray*}
&&\hspace{-1cm} D_z m(y) \\
&&\hspace{-1cm}= m(y + \delta_z ) - m(y) \\
&&\hspace{-1cm}=\int_H L(y,x) \; \textbf{1}_{z \notin \mathcal{C}(y)}[(\lambda-1)+\alpha \dot{x}_z] \mu_X(dx).
\end{eqnarray*}
So
\begin{eqnarray*}
&&\hspace{-1cm}\displaystyle{\nabla_A m(y)}\\
&&\hspace{-1cm}=\displaystyle{ \int_A D_z m(y) \; \nu(dz) }\\
&&\hspace{-1cm}=\displaystyle{\int_{H_S} L(y,x) \int_A \textbf{1}_{z \notin \mathcal{C}(y)}[(\lambda-1)+\alpha \dot{x}_z]\nu(dz) \mu_X(dx)} \\
&&\hspace{-1cm}=\displaystyle{\int_{H_S} L(y,x) \int_A [(\lambda-1)+\alpha \dot{x}_z]\nu(dz)\mu_X(dx), }\\
&&\textrm{ as } \nu \textrm{ is atomless,}\\
&&\hspace{-1cm}=\displaystyle{\int_{H_S} L(y,x)[(\lambda-1)\nu(A) +\alpha x(A)] \mu_X(dx) }\\
&&\hspace{-1cm}=\displaystyle{\int_{H_S} [(\lambda-1)\nu(A)+\alpha x(A)] m(y) \, \mu_{X|Y}(dx\vert y),}\\
&&\textrm{by \textit{iii)}} \, \textrm{ of Proposition \ref{prop:BayesJust}.} 
\end{eqnarray*}
By Proposition \ref{prop:BayesPoisson}, we have that 
$$ \E[X(A)|\mathcal{Y}]=\frac{\nabla_A m(y)}{\alpha \; m(y)}-\frac{(\lambda-1) \nu(A)}{\alpha}, \; A \in \mathcal{B}(S). $$
\end{proof}
\begin{remarks}
\begin{itemize}
\item Neither $\nabla$ nor $D$ satisfy the chain rule of derivation, and consequently $\displaystyle{\frac{\nabla_A F}{F} \neq \nabla_A \log F}$.
\item We have shown in Proposition \ref{rapportBayesGrad} that 
\begin{equation}
\label{BayesPoissonxpoint}
\E[\dot{X}_s|\mathcal{Y}]=\frac{D_s m(Y)}{\alpha \; m(Y)}-\frac{(\lambda-1)}{\alpha}, \; \forall s \in S.
\end{equation}
\end{itemize}
\end{remarks}
\noindent
We conclude this section by a more explicit case, that is the classical Poisson process on a time interval $[0,T]$ equipped with the Lebesgue measure $\pi$. More precisely, let $(X_t)_{t \in [0,T]}$ be an input signal with values in $H_{[0,T]}$ (see Section \ref{section:Analysis on the Poisson space}). The output $(Y_t)_{t \in [0,T]}$ is supposed to be a Poisson process with intensity $\lambda + \alpha X$ where $\lambda$ and $\alpha$ are some fixed parameters. \\
The likelihood denoted by $L$ is given by
\begin{eqnarray*}
&&L(y,x)=\exp \left( -(\lambda-1) T -\alpha \int_S \dot{x}_s \pi(ds) \right)\\ 
&\times&\prod_{k=1}^{y([0,T])} (\lambda+\alpha \, \dot{x}_{z_k}),
\end{eqnarray*}
where $z_k \in \mathcal{C}(y).$\\ 
Proposition \ref{rapportBayesGrad} becomes the following corollary,
\begin{corollary}
\label{corollary:rapportBayesGrad}
Under assumptions of Proposition \ref{rapportBayesGrad} we have that 
\begin{equation}
\label{eq:rapportBayesGradBis}
\E[X_t|\mathcal{Y}]=\frac{\nabla_{[0,t]} m(Y)}{\alpha \, m(Y)}-\frac{(\lambda-1) t}{\alpha}, \; t \in [0,T].
\end{equation}
\end{corollary}

\begin{remark}
\label{remark:application}
The nonlinear filter given by equations (\ref{rapportBayesGrad}) and (\ref{eq:rapportBayesGradBis}) can be numerically approximated by evaluating $m$ in (\ref{eq:m}) by a Monte-Carlo scheme. This computation is really tractable since the Malliavin derivative $\nabla$ is a difference operator.
\end{remark}

\section{Mutual information and conditional mean estimation: the Poisson case}
\label{section:mutualPoisson}
In this section we present the second main result of this paper (Theorem \ref{theorem:mutualPoisson}), \textit{i.e.} the use of relation (\ref{BayesPoissonxpoint}) to recover (in a different manner) a relation between the mutual information of general Poisson channels and the conditional mean estimator of the input which has been established recently in \cite[Theorems 3-4]{GuoShamaiVerduPoisson} (see also \cite{GuoShamaiVerduSA}). We stress that we propose a new proof of this result involving Malliavin calculus and stochastic analysis arguments related to the tools involved in \cite{Zakai} to solve the same problem for additive Gaussian channels. In addition our results are valid for general Poisson channels. Finally we provide extended De Bruijn identities (in Proposition \ref{DeBruijntypeidentities}) of the form of those obtained in \cite[Section VI]{Zakai}. During this section we assume that hypotheses (H1) and (H2) of Section \ref{section:cmePoisson} are in force. First we state and prove the following lemma.  
\begin{lemma}
\label{lemma:gradlogm}
For any $z$ in $\mathcal{S}$ we have that
\begin{eqnarray*}
D_z \log(m)&=&\log\left(1+\frac{D_z m}{m}\right)\\
&=&\log\left(\E[\lambda+\alpha \dot{X}_s\vert \mathcal{Y}]\right), \; \P-a.s. 
\end{eqnarray*}
\end{lemma}
\begin{proof}
First we recall that the last equality follows from relation (\ref{BayesPoissonxpoint}). Then from the definition of the Malliavin derivative $D$ we have that
$$1+\frac{D_z m(y)}{m(y)}=\frac{m(y+\delta_z)}{m(y)}$$
leading to
\begin{eqnarray*}
&&\log\left(1+\frac{D_z m(y)}{m(y)}\right)\\
&=&\log(m(y+\delta_z))-\log(m(y))\\
&=&\log m(y+\delta_z)-\log m(y)\\
&=&D_z (\log m)(y).
\end{eqnarray*}
\end{proof}
\noindent 
In the next Proposition we present extended De Bruijn identities which are the counterpart of \cite[Section VI relation (35)]{Zakai}. These relations will be necessary in the proof of Theorem \ref{theorem:mutualPoisson}. We introduce the following condition 
\begin{equation}
\label{Bayesianriskfinite}
\E\left[\int_S \vert \dot{X}_s \log(\dot{X}_s)\vert \nu(ds)\right]<\infty
\end{equation}
which ensures by Jensen's inequality that the Bayesian risk defined as
$$\left\vert\E\left[\int_S (\dot{X}_s \log(\dot{X}_s) - \E[\dot{X}_s\vert \mathcal{Y}] \log(\E[\dot{X}_s\vert \mathcal{Y}])) \nu(ds) \right]\right\vert$$
is finite.
\begin{prop}[Extended De Bruijn identities]
\label{DeBruijntypeidentities}
Assume that condition (\ref{Bayesianriskfinite}) is satisfied then the relations i) and ii) below hold.
\begin{itemize}
\item[i)]
\begin{eqnarray*}
&&\frac{d}{d\alpha} \E_1[\log m]\\
&&=\frac{1}{\alpha}\E\left[ \int_S \psi_\lambda(\E[\lambda+\alpha \dot{X}_s\vert\mathcal{Y}]) \nu(ds)\right]\\
&&=\frac{1}{\alpha}\E\left[ \int_S \psi_\lambda\left(1+\frac{D_s m}{m}\right) \nu(ds)\right],
\end{eqnarray*}
where $\psi_\lambda(x):=(x-\lambda) \log(x)$ and
\item[ii)] 
\begin{eqnarray*}
&&\frac{d}{d\lambda} \E_1[\log m]\\
&&=\E\left[ \int_S \log\left(\E[\lambda+\alpha \dot{X}_s\vert\mathcal{Y}]\right) \nu(ds)\right]\\
&&=\E\left[ \int_S \log\left(1+\frac{D_s m}{m}\right) \nu(ds)\right]
\end{eqnarray*}
\end{itemize} 
\end{prop}
\begin{proof}
\begin{itemize}
\item[i)]In the following computations we will use the integration by parts formula and the relation 
\begin{equation}
\label{eq:DL}
D_s L(y,x)=L(y,x) (\lambda-1+\alpha\dot{x}_s)
\end{equation}
which has been obtained in the proof of Proposition \ref{rapportBayesGrad}. We recall that $\E_1$ denotes the expectation with respect to $m d\P$ and that $\E_0$ denotes the expectation under $\P$. We follow the main lines of the proof of \cite[Proposition 5.1]{Zakai} with however significant differences like the use of the Malliavin integration by parts formula.
\begin{eqnarray*}
&&\hspace{-1cm}\frac{d}{d\alpha} \E_1[\log m]\\
&&\hspace{-1cm}=\frac{d}{d\alpha} \E_0[m \log m]\\
&&\hspace{-1cm}=\E_0\left[\log m \frac{d m}{d\alpha}\right]+0\\
&&\hspace{-1cm}=\E_0\left[\log m \int_H \left( \frac{d}{d\alpha} \log L(y,x) \right)L(y,x) \mu_X(dx) \right]\\
&&\hspace{-1cm}=\E_0 \bigg[\log m \int_H L(y,x) \bigg( \int_S -\dot{x}_s \nu(ds)\\
&&+\int_0^T \frac{\dot{x}_s}{\lambda+\alpha \dot{x}_s} dy_s \bigg) \mu_X (dx) \bigg]\\
&&\hspace{-1cm}=\int_H \E_0\left[\log m L(y,x) \int_S \frac{-\alpha \dot{x}_s^2+(1-\lambda) \dot{x}_s}{\lambda+\alpha \dot{x}_s} \nu(ds) \right]\\
&&\hspace{-1cm}+\E_0\left[\log m L(y,x) I_1\left(\frac{\dot{x}}{\lambda+\alpha \dot{x}}\right) \right]\mu_X(dx)
\end{eqnarray*}
where we recall that under $\P$ the stochastic process $y$ is a Poisson process with intensity $\nu$ and by definition of the operator $I_1$ we have that 
$$I_1\left(\frac{\dot{x}}{\lambda+\alpha \dot{x}}\right)=\int_S \frac{\dot{x}_s}{\lambda+\alpha \dot{x}_s} (dy_s-\nu(ds))$$
As a consequence:
\begin{eqnarray*}
&&\hspace{-1cm}\frac{d}{d\alpha} \E_1[\log m]\\
&&\hspace{-1cm}=\int_H \E_0\left[\log m L(y,x) \int_S \frac{-\alpha \dot{x}_s^2+(1-\lambda) \dot{x}_s}{\lambda+\alpha \dot{x}_s} \nu(ds) \right] \\
&&\hspace{-1cm}+\E_0\left[\int_S D_s(\log m L(y,x)) \frac{\dot{x}_s}{\lambda+\alpha \dot{x}_s}\nu(ds) \right]\mu_X(dx)
\end{eqnarray*}
where the last equality is obtained using the integration by parts formula (\ref{IBP}). Applying the chain rule formula for the Malliavin derivative (\ref{chainrulePoisson}) we deduce that
\begin{eqnarray*}
&&\hspace{-1cm}\frac{d}{d\alpha} \E_1[\log m]\\
&&\hspace{-1.5cm}=\int_H \E_0\left[\log m L(y,x) \int_S \frac{-\alpha \dot{x}_s^2+(1-\lambda) \dot{x}_s}{\lambda+\alpha \dot{x}_s} \nu(ds) \right]\\
&&\hspace{-1cm}+\E_0\left[\int_S D_s(\log m) L(y,x) \frac{\dot{x}_s}{\lambda+\alpha \dot{x}_s}\nu(ds) \right] \\
&&\hspace{-1cm}+\E_0\left[\int_S D_s(\log m) D_s L(y,x) \frac{\dot{x}_s}{\lambda+\alpha \dot{x}_s}\nu(ds) \right] \\
&&\hspace{-1cm}+\E_0\left[\int_S \log m D_s L(y,x) \frac{\dot{x}_s}{\lambda+\alpha \dot{x}_s}\nu(ds) \right]\mu_X(dx).
\end{eqnarray*}
In addition relation (\ref{eq:DL}) and the preceding expression entail that  
\begin{eqnarray*}
&&\hspace{-1.5cm}\frac{d}{d\alpha} \E_1[\log m]\\
&&\hspace{-1.5cm}=\int_H \E_0\left[\log m L(y,x) \int_S \frac{-\alpha \dot{x}_s^2+(1-\lambda) \dot{x}_s}{\lambda+\alpha \dot{x}_s} \nu(ds) \right]\\
&&\hspace{-1.5cm}+\E_0\left[\int_S D_s(\log m) L(y,x) \frac{\dot{x}_s}{\lambda+\alpha \dot{x}_s}\nu(ds) \right] \\
&&\hspace{-1.5cm}+\E_0\left[\int_S D_s(\log m) L(y,x) \frac{(\lambda-1+\alpha \dot{x}_s)\dot{x}_s}{\lambda+\alpha \dot{x}_s}\nu(ds) \right] \\
&&\hspace{-1.5cm}+\E_0\bigg[\int_S \log m L(y,x) \\
&&\hspace{0cm}\times \frac{(\lambda-1+\alpha\dot{x}_s)\dot{x}_s}{\lambda+\alpha \dot{x}_s}\nu(ds) \bigg]\mu_X(dx) \\
&&\hspace{-1.5cm}=\int_H \E_0\left[\int_S D_s(\log m) L(y,x) \dot{x}_s \nu(ds) \right]\mu_X(dx) \\
&&\hspace{-1.5cm}=\E_0\left[\int_S D_s(\log m) \int_H \dot{x}_s L(y,x) \mu_X(dx) \nu(ds) \right] \\
&&\hspace{-1.5cm}=\E_0\left[\int_S D_s(\log m) m \int_H \dot{x}_s \mu_{X\vert Y}(dx) \nu(ds) \right] \\
&&\hspace{-1.5cm}=\E\left[\int_S D_s(\log m) \E[\dot{X}_s\vert \mathcal{Y}] \right] \\
&&\hspace{-1.5cm}=\frac{1}{\alpha} \E\bigg[\int_S \log \left( 1+\frac{D_s m}{m}\right) \E[\dot{X}_s\vert \mathcal{Y}] \nu(ds) \bigg]\\
&&\hspace{-1.5cm}=\frac{1}{\alpha} \E\left[\int_S \log \left( 1+\frac{D_s m}{m} \right) \left(1+\frac{D_s m}{m}\right) \nu(ds) \right]\\
\end{eqnarray*}
where the last equality comes from Lemma \ref{lemma:gradlogm} and relation (\ref{BayesPoissonxpoint}).
\item[ii)]
The proof is similar to the proof of point i). However for making this paper self-contained we present the main arguments in the following computations.
\begin{eqnarray*}
&&\hspace{-1.5cm}\frac{d}{d\lambda} \E_1[\log m]\\
&&\hspace{-1.5cm}=\frac{d}{d\lambda} \E_0[m \log m]\\
&&\hspace{-1.5cm}=\E_0\left[\log m \frac{d m}{d\lambda}\right]+0\\
&&\hspace{-1.5cm}=\E_0\bigg[\log m \int_H \left( \frac{d}{d\lambda} \log L(y,x) \right)\\
&&\times L(y,x) \mu_X(dx) \bigg]\\
&&\hspace{-1.5cm}=\E_0 \bigg[\log m \int_H L(y,x) \bigg(-\nu(S)\\
&&\hspace{1cm}+\int_S \frac{1}{\lambda+\alpha \dot{x}_s} dy_s \bigg) \mu_X (dx) \bigg]\\
&&\hspace{-1.5cm}=\int_H \E_0\left[\log m L(y,x) \int_S \frac{-\alpha \dot{x}_s-(\lambda-1)}{\lambda+\alpha \dot{x}_s} \nu(ds) \right]\\
&&\hspace{-1.5cm}+\E_0\left[\log m L(y,x) I_1\left(\frac{1}{\lambda+\alpha \dot{x}}\right) \right]\mu_X(dx)\\
&&\hspace{-1.5cm}=\int_H \E_0\left[\log m L(y,x) \int_S \frac{-\alpha \dot{x}_s-(\lambda-1)}{\lambda+\alpha \dot{x}_s} \nu(ds) \right] \\
&&\hspace{-1.5cm}+\E_0\left[\int_S D_s(\log m L(y,x)) \frac{1}{\lambda+\alpha \dot{x}_s}\nu(ds) \right]\mu_X(dx)\\
&&\hspace{-1.5cm}=\int_H \E_0\left[\log m L(y,x) \int_S \frac{-\alpha \dot{x}_s-(\lambda-1)}{\lambda+\alpha \dot{x}_s} \nu(ds) \right]\\
&&\hspace{-1.5cm}+\E_0\left[\int_S D_s(\log m) L(y,x) \frac{1}{\lambda+\alpha \dot{x}_s}\nu(ds) \right] \\
&&\hspace{-1.5cm}+\E_0\left[\int_S D_s(\log m) L(y,x) \frac{\lambda-1+\alpha \dot{x}_s}{\lambda+\alpha \dot{x}_s}\nu(ds) \right] \\
&&\hspace{-1.5cm}+\E_0\left[\int_S \log m L(y,x) \frac{\lambda-1+\alpha\dot{x}_s}{\lambda+\alpha \dot{x}_s}\nu(ds) \right]\mu_X(dx) \\
&&\hspace{-1.5cm}=\int_H \E_0\left[\int_S D_s(\log m) L(y,x) \nu(ds) \right]\mu_X(dx) \\
&&\hspace{-1.5cm}=\E_0\left[\int_S D_s(\log m) m \int_H \mu_{X\vert Y}(dx) \nu(ds) \right] \\
&&\hspace{-1.5cm}=\E\left[\int_S \log \left( 1+\frac{D_s m}{m}\right) \nu(ds) \right]
\end{eqnarray*}
\end{itemize}
\end{proof}
\begin{theorem}
\label{theorem:mutualPoisson}
Assume that condition (\ref{Bayesianriskfinite}) is satisfied then we have that 
\begin{itemize}
\item[i)]
\begin{eqnarray*}
&&\hspace{-1.5cm}\frac{d}{d\alpha} I(X;Y)\\
&&\hspace{-1.5cm}=\frac{1}{\alpha}\E\left[ \int_S \psi_\lambda(\alpha \dot{X}_s+\lambda)-\psi_\lambda(\E[\alpha \dot{X}_s\vert\mathcal{Y}]) \nu(ds)\right]\\
&&\hspace{-1.5cm}=\frac{1}{\alpha}\E\left[ \int_S \psi_\lambda(\alpha \dot{X}_s+\lambda)-\psi_\lambda\left(\frac{D_s m}{m}-(\lambda-1)\right) \nu(ds)\right],
\end{eqnarray*}
where $\psi_\lambda(x):=(x-\lambda) \log (x).$
\item[ii)]
\begin{eqnarray*}
&&\hspace{-1.5cm}\frac{d}{d\lambda} I(X;Y)\\
&&\hspace{-1.5cm}=\E\left[ \int_S \log(\alpha \dot{X}_s+\lambda)-\log(\E[\alpha \dot{X}_s\vert\mathcal{Y}]) \nu(ds)\right]\\
&&\hspace{-1.5cm}=\E\left[ \int_S \log(\alpha \dot{X}_s+\lambda)-\log\left(1+\frac{D_s m}{m}\right) \nu(ds)\right]
\end{eqnarray*}
\end{itemize}
\end{theorem}
\begin{proof}
First we have that
\begin{eqnarray*}
&&I(X;Y)\\
&&=\int_{H\times \Omega} \log\left( \frac{d\mu_{Y\vert X}(\cdot\vert x)}{d \P}(x,y)\right)\\ 
&&\hspace{1cm}-\log\left( \frac{d\mu_Y}{d\mu_W}(y)\right) \mu(dx,dy).\\
&&=\int_H \E_1\left[ \log L(y,x) \right]\mu_X(dx)-\E_1[\log (m)].
\end{eqnarray*}
\begin{itemize}
\item[i)] We have that
\begin{eqnarray}
\label{eqtheotemp1}
&&\hspace{-1cm}\frac{d}{d\alpha} \int_H \E_1\left[ \log L(y,x) \right] \mu_X(dx)\nonumber\\
&&\hspace{-1cm}=\frac{d}{d\alpha} \int_H \E_1\bigg[ -\int_S (\lambda-1)+\alpha\dot{x}_s \nu(ds) \mu_X(dx)\nonumber\\
&&\hspace{1cm}+\int_S \log(\lambda+\alpha \dot{x}_s) dy_s \bigg]\nonumber\\
&&\hspace{-1cm}=\frac{d}{d\alpha} \int_H \int_S -(\lambda-1)-\alpha\dot{x}_s \nonumber\\
&&+\log(\lambda+\alpha \dot{x}_s) (\lambda+\alpha \dot{x}_s)\nu(ds)\mu_X(dx)\nonumber\\
&&\hspace{-1cm}=\int_H \int_S \dot{x}_s \log(\lambda+\alpha \dot{x}_s) \nu(ds) \mu_X(dx)\nonumber\\
&&\hspace{-1cm}=\E\left[ \int_S \dot{x}_s \log(\lambda+\alpha \dot{x}_s) \nu(ds) \mu_X(dx)\right].
\end{eqnarray}
By Proposition \ref{DeBruijntypeidentities} i) it holds that
\begin{eqnarray}
\label{eqtheotemp2}
&&\hspace{-1cm}\frac{d}{d\alpha} \E_1[\log m]\\
&&\hspace{-1cm}=\frac{1}{\alpha}\E\left[ \int_S \log\left(\E[\lambda+\alpha \dot{X}_s\vert\mathcal{Y}]\right) \E[\alpha \dot{X}_s\vert\mathcal{Y}] \nu(ds)\right].\nonumber
\end{eqnarray}
Relations (\ref{eqtheotemp1}) and (\ref{eqtheotemp2}) lead to the result.
\item[ii)]
Similarly we have that
\begin{eqnarray}
\label{eqtheotemp3}
&&\hspace{-1cm}\frac{d}{d\lambda} \int_H \E_1\left[ \log L(y,x) \right] \mu_X(dx)\nonumber\\
&&\hspace{-1cm}=\frac{d}{d\lambda} \int_H \E\bigg[ -\int_S (\lambda-1)+\alpha\dot{x}_s \nu(ds) \mu_X(dx)\nonumber\\
&&+\int_S \log(\lambda+\alpha \dot{x}_s) dy_s \bigg]\nonumber\\
&&\hspace{-1cm}=\frac{d}{d\lambda} \int_H \int_S -(\lambda-1)-\alpha\dot{x}_s\nonumber\\ 
&&+ \log(\lambda+\alpha \dot{x}_s) (\lambda+\alpha \dot{x}_s)\nu(ds)\mu_X(dx)\nonumber\\
&&\hspace{-1cm}=\int_H \int_S \log(\lambda+\alpha \dot{x}_s) \nu(ds) \mu_X(dx)\nonumber\\
&&\hspace{-1cm}=\E\left[ \int_S \log(\lambda+\alpha \dot{x}_s) \nu(ds) \right].
\end{eqnarray}
By Proposition \ref{DeBruijntypeidentities} ii) we have that
\begin{eqnarray*}
\label{eqtheotemp4}
&&\hspace{-2cm}\frac{d}{d\lambda}\E_1[\log m]\nonumber\\
&&\hspace{-2cm}=\frac{1}{\alpha}\E\left[ \int_S \log\left(\E[\lambda+\alpha \dot{X}_s\vert\mathcal{Y}]\right) \nu(ds)\right].
\end{eqnarray*}
We conclude from relations (\ref{eqtheotemp3}) and (\ref{eqtheotemp4}).
\end{itemize}
\end{proof}

\section{A generalization to a class of non-Gaussian and non-Poisson channels}
\label{section:w+P} 
\subsection{The conditional mean estimator formula}
\label{subsection:w+PCME}
In this Section we give a generalization of results from Sections \ref{section:cmePoisson} and \ref{section:mutualPoisson}. We use some notations and definitions presented in Section \ref{section:Appendix}.\\
Let $Y:=(Y_t)_{t\in [0,T]}$ a normal martingale on a probability space $(\Omega,\mathcal{F},\P)$ with a right continuous filtration $(\mathcal{F}_t)_{t\in [0,T]}$ that is:
\begin{itemize}
\item $\E[Y_t^2]<\infty, \; t \in [0,T],$
\item $ \E[Y_t|\mathcal{F}_s]=Y_s, \; 0\leq s<t \leq T$ and
\item $ \E[(Y_t-Y_s)^2|\mathcal{F}_s]=t-s, \; 0 \leq s < t\leq T.$
\end{itemize}
In addition we assume that there exists a predictable function $\phi:=(\phi_t)_{t\in [0,T]}$ such that the stochastic process
\begin{equation}
\label{eq:struct}
Y_t-t-\int_0^t \phi_s \; dY_s, \; t \in [0,T]
\end{equation}
is a martingale. Finally, we assume that $Y$ that has the chaos representation property (see Definition \ref{definition:chaos}). We present two examples of such processes.
\begin{itemize}
\item[Example 1)]
Assume $\phi:=(\phi_t)_{t\in [0,T]}$ appearing in the structure equation (\ref{eq:struct}) is deterministic. Then $(Y_t)_{t\in [0,T]}$ has the chaos representation property see \cite{Emery}, and $(Y_t)_{t\in [0,T]}$ can be represented as
$$ dY_t=i_t dB_t + \phi_t (dN_t-\rho_t dt), \; Y_0=0, \; t \in [0,T] $$
where $(B_t)_{t\in [0,T]}$ is a standard Brownian motion, $i_t=\textbf{1}_{\{\phi_t=0\}}$, $j_t=1-i_t$, and $(N_t)_{t\in [0,T]}$ is a Poisson process independent of $(B_t)_{t\in [0,T]}$ with intensity $\displaystyle{t\mapsto \int_0^t \rho_s ds}$ with $\rho_s:=\frac{j_s}{\phi_s^2}$. Consequently,
\begin{itemize}
\item for $\phi \equiv 1$, $(Y_t)_{t\in [0,T]}$ is a Poisson process with intensity $\displaystyle{\nu_t=\int_0^t \frac{1}{\phi_s^2}\; ds}$;
\item for $\phi \equiv 0$, $(Y_t)_{t\in [0,T]}$ is a standard Brownian motion.
\end{itemize}
\item[Example 2)]
Consider $\phi_t=\beta Y_t, \; \beta \in [-2,0)$. Then $(Y_t)_{t\in [0,T]}$ is an Az\'ema martingale. This process has the chaos decomposition property but its increments are not independent contrary to the previous example.
\end{itemize}
In this Section we assume that assumptions of the Subsection \ref{subsection:generalBayesianfacts} are in force and we recall that we denote by $\E$ the expectation with respect to the measure $\mu_X\times \mu_Y$, $\E_1$ the expectation with respect to $\mu_Y$ and $\E_0$ the expectation relative to $\P$.\\\\ 
Let $\lambda$ and $\alpha$ two positive numbers. Let $(X_t)_{t \in [0,T]}$ a real-valued input process with $\displaystyle{X_t=\int_0^t \dot{X}_s \; ds, \; t \in [0,T]}$. Assume the output signal $(Y_t)_{t \in [0,T]}$ is a normal martingale such that the measure $\mu_{Y|X}(\cdot \vert x)$ is absolutely continuous with respect to $\P$ with likelihood given by 
\begin{eqnarray}
\label{eq:likelihoodsection4}
&&\hspace{-2cm}L(y,x)\nonumber\\
&&\hspace{-2cm}=\frac{d \mu_{Y|X}(\cdot \vert x)}{d \P}\nonumber\\
&&\hspace{-2cm}=\exp \bigg( \int_0^T \lambda-1+\alpha \dot{x}_s dy_s\nonumber\\
&&\hspace{-2cm}-\frac12 \int_0^T (\lambda-1+\alpha\dot{x}_s)^2 \textbf{1}_{\{\phi_s=0\}} \; ds \bigg) \nonumber\\
&&\times\prod_{s\leq T} (\lambda+\alpha\dot{x}_s \phi(s)) e^{-(\lambda-1+\alpha \dot{x}_s) \phi(s)}.\nonumber
\end{eqnarray}
We refer to \cite[Theorem 37, p.~84]{Protter} for technical justifications about the existence of $L$.
\begin{lemma}
With notations of Definition \ref{definition:intstochan} we have,
$$L(y,x)=\sum_{n=0}^\infty \frac{1}{n!} I_n((\lambda-1+\alpha \dot{x})^{\otimes n}),$$
where $(\lambda-1+\alpha \dot{x})^{\otimes n}:[0,T]^n \to \mathbb{R}$ is defined as
$$ (\lambda-1+\alpha \dot{x})^{\otimes n}(t_1,\ldots,t_n)=\prod_{k=1}^n (\lambda-1+\alpha \dot{x}_{t_k}).$$
\end{lemma}
\begin{proof}
See \cite[Section 3.5, p.~87]{Privault}.
\end{proof}
This formulation of $L$ and the definition (\ref{definition:MalliavinMN}) of the Malliavin derivative in this context give
\begin{equation}
\label{D_tl}
D_t L(y,x) =(\lambda-1+\alpha\dot{x}_t) L(y,x), \; t \in [0,T].
\end{equation}
\begin{definition}
For $t$ in $[0,T]$ define $\nabla_t$ as 
$$ \nabla_t F =\int_0^t D_s F \; ds, \; F \in L^2(\Omega) \textrm{ satisfying } (\ref{Dom(D)}).$$
\end{definition}
By using the general Bayesian results presented in Section \ref{section:cmePoisson} we have the following Proposition.
\begin{prop}
$\displaystyle{\E[X_t|\mathcal{Y}]=\frac{\nabla_t m(Y)}{m(Y)}, \; t \in [0,T]}$.
\end{prop} 
\begin{proof}
One can mimic the proof of Proposition \ref{prop:BayesPoisson} by noticing that the key ingredient is formula (\ref{D_tl}). 
\end{proof}

\subsection{Mutual information and conditional mean estimation}
\label{section:mutualmixture}
In this section we consider a particular example of mixtures of Gaussian-Poisson presented in of Section \ref{section:w+P}. Let $(\phi_t)_{[0,T]}$ be a deterministic function with values in $\{0,1\}$ and let $(Y_t)_{t\in [0,T]}$ be the martingale defined on a probability space $(\Omega,\mathcal{F},\P)$ by
$$ dY_t=\textbf{1}_{\phi_t=0} dB_t+\phi_t (dN_t-\pi(dt)),\quad t \in [0,T]$$
where $B$ and $N$ denote respectively a standard Brownian motion and an independent Poisson process with intensity the Lebesgue measure on $[0,T]$ denoted by $\pi$. This model is really an "hand-made" example of a mixture between Gaussian and Poisson regimes. Actually $\phi$ can be thought as a "switch" enabling a user to pass from the Gaussian regime ($\phi_s=0$) to the Poisson one ($\phi_s=1$). In addition please note that we assume no restrictions on the number of switches from one state to another. The next Lemma and Theorem are the main result of this section. 
\begin{lemma}
Assume that 
\begin{equation}
\label{condition2}
\E\left[ \int_0^T \dot{X}_s^2 \textbf{1}_{\phi_s=0} + \vert \dot{X}_s \log(\dot{X}_s) \vert \textbf{1}_{\phi_s=1} ds \right]<\infty
\end{equation}
then the following relations hold
\begin{itemize}
\item[i)]
\begin{eqnarray*}
&&\hspace{-1cm}\frac{d}{d\alpha} \E_1[\log m]\\
&&\hspace{-1cm}=\E\left[\int_0^T \E[(\lambda-1+\alpha \dot{X}_s) \vert \mathcal{Y}] \E[\dot{X}_s \vert \mathcal{Y}] \textbf{1}_{\phi_s=0} ds \right]\\
&&\hspace{-1cm}+\frac{1}{\alpha}\E\left[\int_0^T \log\E[\lambda+\alpha \dot{X}_s \vert \mathcal{Y}] \E[\alpha \dot{X}_s\vert \mathcal{Y}] \textbf{1}_{\phi_s=1} ds \right].
\end{eqnarray*}
\item[ii)]
\begin{eqnarray*}
&&\hspace{-1cm}\frac{d}{d\lambda} \E_1[\log m]\\
&&\hspace{-1cm}=\E\left[\int_0^T \E[(\lambda-1+\alpha \dot{X}_s) \vert \mathcal{Y}] \textbf{1}_{\phi_s=0} ds \right]\\
&&\hspace{-1cm}+\E\left[\int_0^T \log\E[\lambda+\alpha \dot{X}_s \vert \mathcal{Y}] \textbf{1}_{\phi_s=1} ds \right].
\end{eqnarray*}
\end{itemize}
\end{lemma}
\begin{proof}
We only present the proof of i): the one of ii) being very similar. We have that
\begin{eqnarray*}
&&\hspace{-1.5cm}\frac{d}{d\alpha} \E_1[\log m]\\
&&\hspace{-1.5cm}=\E_0\left[\log m \int_H \left( \frac{d}{d\alpha} \log L(y,x) \right)L(y,x) \mu_X(dx) \right]\\
&&\hspace{-1.5cm}=\int_H \E_0\left[\log m L(y,x) I_1\left(\frac{\dot{x}}{1+(\lambda-1+\alpha \dot{x})\phi}\right) \right]\\
&&\hspace{-1.5cm}-\E_0\left[\log m L(y,x) \int_0^T \frac{(\lambda-1+\alpha \dot{x}_s)\dot{x}_s\phi_s}{1+(\lambda-1+\alpha \dot{x}_s)\phi_s} ds \right]\\
&&\hspace{-1.5cm}-\E_0\left[\log m L(y,x) \int_0^T (\lambda-1+\alpha \dot{x}_s)\dot{x}_s \textbf{1}_{\phi_s=0} ds \right]\mu_X(dx).
\end{eqnarray*}
Note that in this situation 
\begin{eqnarray*}
&&\hspace{-1.5cm}I_1\left(\frac{\dot{x}}{1+(\lambda-1+\alpha \dot{x})\phi}\right)\\
&&\hspace{-1.5cm}=\int_0^T \frac{\dot{x}_s}{1+(\lambda-1+\alpha \dot{x}_s)\phi_s} dy_s\\
&&\hspace{-1.5cm}=\int_0^T \frac{\dot{x}_s}{1+(\lambda-1+\alpha \dot{x}_s)\phi_s} (\textbf{1}_{\phi_s=0} dB_s+\phi_s (dN_s-ds)).  
\end{eqnarray*}
Then we make use of the Malliavin integration by parts formula (\ref{IBPmixture}).
\begin{eqnarray*}
&&\hspace{-1.5cm}\frac{d}{d\alpha} \E_1[\log m]\\
&&\hspace{-0.5cm}=\int_H \E_0\left[\int_0^T \frac{\dot{x}_s \tilde{D}_s(\log m L(y,x))}{1+(\lambda-1+\alpha \dot{x}_s)\phi} ds \right]\\
&&\hspace{-0.5cm}-\E_0\left[\log m L(y,x) \int_0^T \frac{(\lambda-1+\alpha \dot{x}_s)\dot{x}_s\phi_s}{1+(\lambda-1+\alpha \dot{x}_s)\phi_s} ds \right]\\
&&\hspace{-0.5cm}- \E_0\left[\log m L(y,x) \int_0^T (\lambda-1+\alpha \dot{x}_s)\dot{x}_s \textbf{1}_{\phi_s=0} ds \right]\hspace{-0.1cm}\mu_X(dx)
\end{eqnarray*}
We need a chain rule formula for the Malliavin derivative $\tilde{D}$ which can be found for example in \cite{Privault}: 
\begin{eqnarray}
\label{chainrulemixture}
&&\tilde{D}_s (\log m L(y,x))\\\nonumber 
&&= L(y,x) \tilde{D}_s (\log m) + \log m \tilde{D}_s L(y,x)\\\nonumber
&&+\phi_s \tilde{D}_s (\log m) \tilde{D}_s (L(y,x)).
\end{eqnarray} 
Combining (\ref{chainrulemixture}) and (\ref{D_tl}) we obtain that 
\begin{eqnarray*}
&&\hspace{-0.5cm}\frac{d}{d\alpha} \E_1[\log m]\\
&&\hspace{-0.5cm}=\int_H \E_0\left[\int_0^T L(y,x) \frac{\dot{x}_s \tilde{D}_s(\log m)}{1+(\lambda-1+\alpha \dot{x}_s)\phi} ds \right]\\
&&\hspace{-0.5cm}+\E_0\left[\int_0^T \frac{\dot{x}_s \tilde{D}_s(\log m) \tilde{D}_s L(y,x) \phi_s}{1+(\lambda-1+\alpha \dot{x}_s)\phi} ds \right]\\
&&\hspace{-0.5cm}+\E_0\left[\int_0^T \log m L(y,x) \frac{\dot{x}_s \tilde{D}_s}{1+(\lambda-1+\alpha \dot{x}_s)\phi} ds \right]\\
&&\hspace{-0.5cm}-\E_0\left[\log m L(y,x) \int_0^T \frac{(\lambda-1+\alpha \dot{x}_s)\dot{x}_s\phi_s}{1+(\lambda-1+\alpha \dot{x}_s)\phi_s} ds \right]\\
&&\hspace{-0.5cm}- \E_0\left[\log m L(y,x) \int_0^T (\lambda-1+\alpha \dot{x}_s)\dot{x}_s \textbf{1}_{\phi_s=0} ds \right]\hspace{-0.1cm}\mu_X(dx)\\
&&\hspace{-0.5cm}=\int_H \E_0\left[\int_0^T L(y,x) \frac{\dot{x}_s \tilde{D}_s(\log m)}{1+(\lambda-1+\alpha \dot{x}_s)\phi} ds \right]\\
&&\hspace{-0.5cm}+\E_0\left[\int_0^T L(y,x) \frac{\tilde{D}_s(\log m) (\lambda-1+\alpha \dot{x}_s)\dot{x}_s \phi_s}{1+(\lambda-1+\alpha \dot{x}_s)\phi} ds \right]\\
&&\hspace{-0.5cm}+\E_0\left[\int_0^T \log m L(y,x) \frac{\dot{x}_s(\lambda-1+\alpha \dot{x}_s)}{1+(\lambda-1+\alpha \dot{x}_s)\phi} ds \right]\\
&&\hspace{-0.5cm}-\E_0\left[\log m L(y,x) \int_0^T \frac{(\lambda-1+\alpha \dot{x}_s)\dot{x}_s\phi_s}{1+(\lambda-1+\alpha \dot{x}_s)\phi_s} ds \right]\\
&&\hspace{-0.5cm}-\E_0\left[\log m L(y,x) \int_0^T (\lambda-1+\alpha \dot{x}_s)\dot{x}_s \textbf{1}_{\phi_s=0} ds \right]\mu_X(dx)\\
&&\hspace{-0.5cm}=\int_H \E_0\left[\int_0^T \tilde{D}_s(\log m) L(y,x) \dot{x}_s ds \right].
\end{eqnarray*}
Since $Y$ is a mixture of Gaussian and Poisson processes one can show that the Malliavin derivative $\tilde{D}$ can be decomposed in two parts $D^B$ and $D$ where $D^B$ acts on the Gaussian (Brownian) part of a functional of $Y$ and where $D$ acts on the Poisson part of it. Actually $D^B$ is related to the Malliavin derivative presented in \cite{Zakai} and $D$ is the difference operator used in sections \ref{section:Analysis on the Poisson space}-\ref{section:mutualPoisson}. More precisely we have that
$$ \tilde{D}_s \log(m)=D_s^B \log(m) \textbf{1}_{\phi_s=0} + \textbf{1}_{\phi_s=1} D_s \log(m).$$
From relation \cite[(19)]{Zakai} and Lemma \ref{lemma:gradlogm} we deduce that
\begin{eqnarray*}
&&\tilde{D}_s \log(m)\\
&&=(\lambda-1+\alpha \dot{x}_s) \textbf{1}_{\phi_s=0} + \textbf{1}_{\phi_s=1}\log\left(1+\frac{D_s m}{m}\right)
\end{eqnarray*}  
leading to
\begin{eqnarray*}
&&\hspace{-0.8cm}\frac{d}{d\alpha} \E_1[\log m]\\
&&\hspace{-0.8cm}=\int_H \E_0\left[\int_0^T \frac{D_s^B m}{m} \textbf{1}_{\phi_s=0} L(y,x) \dot{x}_s ds \right]\\
&&\hspace{-1cm}+\E_0\left[\int_0^T \log\left(1+\frac{D_s m}{m}\right) L(y,x)  \textbf{1}_{\phi_s=1} \dot{x}_s ds \right]\mu_X(dx)\\
&&\hspace{-0.8cm}=\E\left[\int_0^T \E[(\lambda-1+\alpha \dot{X}_s) \vert \mathcal{Y}] \E[\dot{x}_s\vert \mathcal{Y}] \textbf{1}_{\phi_s=0} ds \right]\\
&&\hspace{-0.8cm}+\frac{1}{\alpha}\E\left[\int_0^T \log\E[\lambda+\alpha \dot{X}_s \vert \mathcal{Y}] \E[\alpha \dot{x}_s\vert \mathcal{Y}] \textbf{1}_{\phi_s=1} ds \right].
\end{eqnarray*}
\end{proof}
We conclude this Section by the counterpart of Theorem \ref{theorem:mutualPoisson} in this context.
\begin{theorem}
\label{theorem:mutualW+P}
Assume that condition (\ref{condition2}) is satisfied then we have that
\begin{itemize}
\item[i)]
\begin{eqnarray*}
&&\hspace{-2cm}\frac{d}{d\alpha} I(X;Y)\\
&&\hspace{-2cm}=\frac{1}{\alpha}\E\bigg[ \int_0^T \big[(\lambda-1+\alpha \dot{X}_s) \dot{X}_s\\
&&\hspace{-2cm}-(\E[(\lambda-1+\alpha \dot{X}_s)\vert\mathcal{Y}]) \big] \E[\dot{X}_s\vert\mathcal{Y}]) \big]\textbf{1}_{\phi_s=0} ds\bigg]\\
&&\hspace{-2cm}+\frac{1}{\alpha}\E\left[ \int_0^T \left[\psi_\lambda(\alpha \dot{X}_s+\lambda)-\psi_\lambda(\E[\alpha \dot{X}_s\vert\mathcal{Y}])\right] \textbf{1}_{\phi_s=1} ds\right],
\end{eqnarray*}
where $\psi_\lambda(x):=(x-\lambda) \log (x).$
\item[ii)]
\begin{eqnarray*}
&&\hspace{-2cm}\frac{d}{d\lambda} I(X;Y)\\
&&\hspace{-1cm}=\E\bigg[ \int_0^T \big[(\lambda-1+\alpha \dot{X}_s)\\
&&\hspace{-1cm}-(\E[(\lambda-1+\alpha \dot{X}_s)\vert\mathcal{Y}]) \big] \textbf{1}_{\phi_s=0} ds\bigg]\\
&&\hspace{-1cm}+\E\left[ \int_0^T \left[\log(\alpha \dot{X}_s+\lambda)-\log(\E[\alpha \dot{X}_s\vert\mathcal{Y}])\right] \textbf{1}_{\phi_s=1} ds\right],
\end{eqnarray*}
\end{itemize}
\end{theorem}
\begin{proof}
The proof is very similar to the proof of Theorem \ref{theorem:mutualPoisson}. We just mention that
\begin{eqnarray*}
&&\int_H \E_1[\log L(y,x)] \mu_X(dx)\\
&&=\int_H-\int_0^T (\lambda-1+\alpha \dot{x}_s) \phi_s ds\\
&&+\int_0^T \log(1+(\lambda-1+\alpha \dot{x}_s)\phi_s)(\lambda+\alpha \dot{x}_s) ds\\
&&-\frac12 \int_0^T (\lambda-1+\alpha \dot{x}_s)^2 \textbf{1}_{\phi_s=0} ds \mu_X(dx).
\end{eqnarray*}
\end{proof}
\begin{remark}
\begin{itemize}
\item We recover the result of \cite{GuoShamaiVerduGaussian} and \cite{Zakai} by taking $\phi \equiv 0$ and $\lambda=1$ in Theorem \ref{theorem:mutualW+P} i). Note also that when $\phi \equiv 0$ the case $\lambda\neq 1$ is a bit artificial since we know that coefficients $\lambda$ and $\alpha$ can be replaced by a single parameter: the signal to noise ratio (which coincides with $\alpha$ when $\lambda=1$). 
\item We recover the result of \cite{GuoShamaiVerduPoisson} and Theorem \ref{theorem:mutualPoisson} by taking $\phi \equiv 1$ in Theorem \ref{theorem:mutualW+P}.   
\end{itemize}
\end{remark}

\section{Appendix}
\label{section:Appendix}
In this Appendix we give some further elements of stochastic analysis in the framework of normal martingales. We use notations of Section \ref{section:w+P}. 

\begin{definition}
\label{definition:intstochan}
Let $Y$ be a normal martingale. For $n \geq 1$, let $L^2([0,T])^{\circ n}$ be the space of symmetric functions $f_n$ in $n$ variables. For, $f_n$ in $L^2([0,T])^{\circ n}$ define the iterated stochastic integral $I_n(f_n)$ by
\begin{eqnarray*}
&&\disp{I_n(f_n)}\\
&:=&\disp{n! \int_0^T \int_0^{t_n} \ldots \int_0^{t_2} f_n(t_1,\cdots,t_n) \, dY_{t_1} \ldots dY_{t_n}.}
\end{eqnarray*}
For $f_0$ in $\real$ we let $I_0(f_0):=f_0$.  
\end{definition} 
\noindent
In addition we have that
$$ I_n(f_n) = n \int_0^T I_{n-1}(f_n(\ast,t)\textbf{1}_{[0,t]^{n-1}}(\ast)) \, dY_t, \; n \geq 1,$$
where $f_n(\ast,t)$ denotes the elements in $L^2([0,T])^{\circ n-1}$ obtained by considering $f_n$ where one variable is fixed at $t$ (since $f_n$ is symmetric we assume that the first variable is fixed to be equal to $t$). 
\begin{definition}
\label{definition:chaos}
\item Denote for $n \geq 1$, 
$$ \mathcal{H}_n = \{ I_n(f_n), \; f_n \in L^2([0,T])^{\circ n} \}. $$
\item We say that $(Y_t)_{t \in [0,T]}$ has the chaos representation property if 
$$ L^2(\Omega) = \bigoplus_{n=0}^{\infty} \mathcal{H}_n,$$
that is, for every $F$ in $L^2(\Omega)$ there exists $(f_n)_{n \in \inte}$ such that $f_n \in L^2([0,T])^{\circ n}, \; n \geq 1$ and  
$$ F=\sum_{n=0}^{\infty} I_n(f_n).$$
\end{definition}
\noindent
As an example this property is true for the mixture of Gaussian and Poisson processes considered in Section \ref{section:w+P}.\\\\
\noindent
We introduce the Malliavin derivative with respect to $(Y_t)_{t \in [0,T]}$. Let 
\begin{eqnarray*}
&&\mathcal{S}=\bigg\{ \sum_{k=0}^n I_k(f_k), \; f_k \in L^2([0,T])^{\circ k}, \\
&&\hspace{1.5cm} 0\leq k\leq n, n \in \inte \bigg\}.
\end{eqnarray*}
We define the Malliavin derivative $D$ as the linear operator from $\mathcal{S}$ to $L^2(\Omega \times [0,T])$ by
\begin{equation}
\label{definition:MalliavinMN}
D_t I_n(f_n) = n I_{n-1}(f_n(\ast,t)), \; d\P \times dt-a.e. 
\end{equation}
We state the Malliavin integration by parts relative to the process $Y$. Let $F$ in $L^2(\Omega)$ and denote, as above, by $(f_n)_n$ the functions appearing in its chaotic decomposition. Assume that 
\begin{equation}
\label{Dom(D)}
\sum_{n=1}^\infty n n! \|f_n\|_{L^2([0,T]^n)}^2<\infty
\end{equation}
then for every deterministic $h:[0,T]\to \mathbb{R}$ we have that
\begin{equation}
\label{IBPmixture}
\E[F I_1(h)]=\E\left[\int_0^T \tilde{D}_s F h(s) ds\right]
\end{equation}
where $\E$ denotes the expectation relative to $\P$.

\section*{Acknowledgment}
I am grateful to two anonymous referees whose comments and suggestions have led to major improvements of this paper.

\end{document}